\documentclass[twocolumn,secnumarabic,amssymb, nobibnotes, aps, prl,superscriptaddress]{revtex4-2}

\usepackage{physics}
\usepackage{mathtools} 
\usepackage{comment} 
\usepackage{graphicx}
\usepackage[hidelinks]{hyperref}
\usepackage{xcolor}
\usepackage{dcolumn}
\usepackage{threeparttable}

\begin{document}

\title{Davies equation without the secular approximation:\\ Reconciling locality with quantum thermodynamics for open quadratic systems}%

\author{Koki Shiraishi}%
\email{shiraishi@cat.phys.s.u-tokyo.ac.jp}
\affiliation{
  Department of Physics, The University of Tokyo, 7-3-1 Hongo, Bunkyo-ku, Tokyo 113-0033, Japan.
}
\author{Masaya Nakagawa}%
\affiliation{
  Department of Physics, The University of Tokyo, 7-3-1 Hongo, Bunkyo-ku, Tokyo 113-0033, Japan.
}
\author{Takashi Mori}%
\affiliation{
  Department of Physics, Keio University, Kohoku-ku, Yokohama, Kanagawa 223-8522, Japan.
}
\date{July 2025}%
\begin{abstract}
We derive a thermodynamically consistent quantum master equation that satisfies locality for quadratic systems coupled to independent and identical baths at each site.
We show that the quasi-local Redfield equation coincides exactly with the Davies equation, which satisfies the detailed-balance condition, due to cancellation of quantum coherence generated by each bath.
This derivation does not rely on the secular approximation, which fails in systems with vanishing energy-level spacings.
We discuss generalizations of our result to slowly driven quadratic systems and generic quantum many-body systems. Our result paves the way to a thermodynamically consistent description of quantum many-body systems.
\end{abstract}
\maketitle

\paragraph{Introduction.}
The detailed balance condition, which means that every transition is balanced with its reverse in equilibrium, lies at the heart of statistical mechanics and stochastic thermodynamics.
It first appeared in Boltzmann's H-theorem~\cite{Boltzmann1872} and later in Einstein's relations for absorption and emission~\cite{Einstein1917}.
Since the 1970's, the detailed balance condition has been generalized to open quantum dynamics by requiring the generator to respect the symmetry properties of the thermal equilibrium state~\cite{Agarwal1973,Alicki1976,Kossakowski1977,10.1002/9780470142578.ch2, 10.1063/1.526164,doi:10.1142/S0219025707002762,Fagnola2008,fagnola2009two,PhysRevA.106.062209,scandi2025thermalizationopenmanybodysystems}.
Davies' seminal work~\cite{Davies1974,Davies1976} showed that, for a system weakly coupled to a thermal reservoir, the Born-Markov and secular approximations produce a quantum master equation (QME) that rigorously satisfies the detailed balance condition~\cite{Alicki1976,10.1002/9780470142578.ch2,Fagnola2008,fagnola2009two}.  
The Davies equation ensures relaxation to the Gibbs state, non-negative entropy production~\cite{10.1063/1.523789}, and fluctuation theorems~\cite{PhysRevResearch.4.023230,PhysRevA.106.062209}, and has versatile applications to quantum thermodynamics~\cite{RevModPhys.81.1665,Alicki2018,doi:10.1063/1.5096173,10.1063/1.5115323}.

However, in quantum many-body systems, making QMEs consistent with thermodynamics has long been a major challenge, because the Davies equation faces two  serious difficulties~\cite{PhysRevE.76.031115,PhysRevA.101.012103,mori_annurev-conmatphys-040721-015537,stefanini2025lindbladme}.  
The first difficulty is the nonlocality of the transitions between energy eigenstates.  
When the bath and the system interact locally, the QME describing the time evolution of the system should also reflect its quasi-locality.
However, in the Davies equation, the effect of the bath is described as transitions between energy eigenstates, which are generally highly nonlocal.  
The second difficulty is the breakdown of the secular approximation.  
The secular approximation assumes that energy-level spacings are sufficiently large; as the system size increases, the energy-level spacings become small and this approximation fails.

Since Levy and Kosloff~\cite{Levy_2014} first pointed out violations of the second law of thermodynamics in local QMEs, there have been extensive efforts to reconcile locality with thermodynamic consistency~\cite{Trushechkin_2016,https://doi.org/10.1002/prop.201600067, Hofer_2017,doi:10.1142/S1230161217400108,doi:10.1142/S1230161217400108,DeChiara_2018,Cattaneo_2019,PhysRevResearch.3.013165,Potts_2021,PhysRevResearch.4.023230,PhysRevA.105.032208,PhysRevE.107.014108,PhysRevA.107.062216}.  
Despite these advances, a general framework that simultaneously guarantees both locality of the QMEs and strict thermodynamic consistency remains elusive. 
To avoid the breakdown of the secular approximation, microscopic derivations using partial-secular approximations~\cite{doi:10.1063/1.4907370,https://doi.org/10.48550/arxiv.1710.09939,PhysRevA.100.012107}, time-coarse-graining~\cite{Cohen-Tannoudji_1986,LIDAR200135,PhysRevA.78.022106,PhysRevA.79.032110,PhysRevA.88.012103,Mozgunov2020completelypositive}, and various other schemes~\cite{PhysRevB.102.115109,PhysRevB.97.035432,Davidovic2020completelypositive} have been proposed, but their thermodynamic consistency remains unclear.  
More recently, a microscopic derivation of the QME satisfying quasi-locality and a relaxed form of detailed balance has been proposed~\cite{scandi2025thermalizationopenmanybodysystems}, yet the derivation of the Davies equation satisfying a stronger form of the detailed balance has not been done without the secular approximation.

In this Letter, we consider a quantum many-particle system with a quadratic Hamiltonian, in which each site is coupled to identical independent baths, and show that the Redfield equation~\cite{5392713,REDFIELD19651} obtained under the Born-Markov approximation coincides exactly with the Davies equation without any further approximations.  
Since the Redfield equation satisfies quasi-locality~\cite{PhysRevB.111.184311,1q5b-p2bf}, this equivalence demonstrates the coexistence of quasi-locality and the thermodynamic laws embodied by the Davies equation.  
Moreover, because we employ no approximations beyond the Born-Markov approximation, our derivation holds regardless of the breakdown of the secular approximation.
Although our result is limited to quadratic systems and does not directly apply to general interacting many-body systems, the system we consider has broad applicability in studies of non-equilibrium steady states, transport, and entanglement dynamics~\cite{Prosen_2008, Prosen_2010, PTProsen_2010PT, PhysRevLett.107.137201, caceffo2024fateentanglementquadraticmarkovian}.  
We also discuss possible extensions of our results to a more general type of dissipation and many-body systems.

\paragraph{Model.}
We consider an $N$-site quadratic fermion (or boson) system connected to free-fermion (or boson) baths at each site.
Each bath is assumed to be in thermal equilibrium. 
The total Hamiltonian of the system and baths is given as $H_{\mathrm{tot}}=H_{\mathrm{S}}+\sum_j(H_{\mathrm{B},j}+H_{\mathrm{SB},j})$, where $j$ is a site index.
The site index may include internal degrees of freedom such as spin and sublattice.
We assume that the Hamiltonians of the system and the baths separately conserve the particle number and that the system-bath interaction conserves the total number of particles.
The Hamiltonian of the system $H_{\mathrm{S}}$ is generally written as
\begin{equation}
  \label{eq:H_system}
  H_\mathrm{S}=\sum_{j,k=1}^Nh_{jk}a_j^\dagger a_k=\sum_{m=1}^N\omega_m c_m^\dagger c_m,
\end{equation}
where $a_j$ and $a_j^\dag$ are the annihilation and creation operators of a particle at site $j$, respectively, and $c_m$ and $c_m^\dagger$ are those of a particle in energy eigenmode $m$.
They are related as
\begin{equation}
  \label{eq:adecomp}
  a_j=\sum_{m}V_{mj}c_m,~a_j^\dagger=\sum_{m}V_{mj}^*c_m^\dagger,
\end{equation}
where $V$ is a unitary matrix.
The Hamiltonian $H_{\mathrm{B},j}$ of the bath and the system-bath interaction Hamiltonian $H_{\mathrm{SB},j}$ at site $j$ are given as
\begin{align}
  \label{eq:H_bath}
  H_{\mathrm{B},j}&=\int dk\, \omega(k) d_k^{(j)\dag} d_k^{(j)},\\\label{eq:interaction}
  H_{\mathrm{SB},j}&=J_{\mathrm{int},j}a_j^\dag\sqrt{\frac{1}{(2\pi)^D}}\int dk\, d_k^{(j)}+\mathrm{H.c.},
\end{align}
where $d_k^{(j)}$ and $d_k^{(j)\dag}$ are the annihilation and creation operators of a particle in mode $k$ in a bath at site $j$, $\omega(k)$ is the dispersion relation of the bath, $J_{\mathrm{int},j}\geq0$ is the coupling strength, and $D$ is the spatial dimension of the bath.
We assume that all the baths are identical and have the same dispersion relation. 

The state $\rho_{B,j}$ of the bath at site $j$ is given by the Gibbs state $\rho_{\mathrm{B},j}\propto\exp[-\beta (H_{\mathrm{B},j}-\mu\int dk d_k^{(j)\dag} d_k ^{(j)})]$,
where we assume that all baths have the same inverse temperature $\beta$ and the chemical potential $\mu$.
The correlation function of the bath at site $j$ is defined as $C_{\mu\nu}^{(j)}(t)=\tr[B_{\mu,j}^\dagger(t)B_{\nu,j}\rho_{\mathrm{B},j}]$~($\mu,\nu=1,2$),
where the operator $B_{\mu,j}$ of the bath is given as
\begin{equation}
  B_{1,j}\coloneqq \sqrt{\frac{1}{(2\pi)^D}}\int dk d_k^{(j)},~B_{2,j}\coloneqq \sqrt{\frac{1}{(2\pi)^D}}\int dk d_k^{(j)\dag}.
\end{equation}
Since all the baths are identical, the correlation functions do not depend on $j$: $C_{\mu\nu}^{(j)}(t)=C_{\mu\nu}(t)$.
The one-sided Fourier transform of the correlation function $C_{\mu\nu}(t)$ is expressed as
\begin{equation}
  \label{eq:defgammaeta}
  \Gamma_{\mu\nu}(\omega)\coloneqq\int_{0}^\infty{ds}e^{i\omega s}C_{\mu\nu}(s)=\frac{1}{2}\gamma_{\mu\nu}(\omega)+i\eta_{\mu\nu}(\omega),
\end{equation}
where $\gamma_{\mu\nu}(\omega)$ is the power spectrum function, which is given by the Fourier transform of the correlation function $C_{\mu\nu}(t)$, and $\eta_{\mu\nu}(\omega)$ is related to $\gamma_{\mu\nu}(\omega)$ via the Kramars-Kronig relation.
Specifically, the power spectrum functions are given as
\begin{equation}
  \label{eq:gamma_bath}
  \begin{split}
    \gamma_{11}(\omega)&=(1\mp f_{\beta,\mu,\pm}(\omega))D(\omega),\\
    \gamma_{22}(-\omega)&=f_{\beta,\mu,\pm}(\omega)D(\omega),\\
    \gamma_{12}(\omega)&=\gamma_{21}(\omega)=0,
  \end{split}
\end{equation}
where $D(\omega)$ is the density of states of the baths and $f_{\beta,\mu,\pm}(\omega)=1/(e^{\beta(\omega-\mu)}\pm 1)$ is the Fermi (Bose) distribution function for fermionic (bosonic) baths.
They satisfy the relation
\begin{equation}
  \label{eq:KMScond}
\gamma_{11}(\omega)=e^{\beta(\omega-\mu)}\gamma_{22}(-\omega),
\end{equation}
which corresponds to the Kubo-Martin-Schwinger (KMS) condition.

\paragraph{Redfield equation and Davies equation.} 
A standard derivation with the Born-Markov approximation gives the Redfield equation, which is the time-evolution equation of the density matrix of the system~\cite{5392713,REDFIELD19651} (here we set $\hbar=1$):
\begin{equation}
  \label{eq:RedfieldSchr}
  \begin{split}
    &\frac{d}{dt}\rho(t)=-i[H_\mathrm{S},\rho(t)]\\
    &+\sum_j\sum_{\mu,\nu}\int_0^\infty{ds}\left[ J_{\mathrm{int},j}^2C_{\mu\nu}(s)(A_\nu(-s)\rho(t)A_\mu^{\dagger}\right.\\
    &~~~~~~~~~~~~~~~~~~~~\left.-A_\mu^{\dagger} A_\nu(-s)\rho(t))+\mathrm{H.c.}\right],
  \end{split}
\end{equation}
where $A_1=a_j^\dag$ and $A_2=a_j$.
The interaction representation of the operator $A_\mu$ is written as $A_\mu(s)=\exp(isH_{\mathrm{S}})A_\mu\exp(-isH_{\mathrm{S}})$.
Importantly, the Redfield equation (9) satisfies locality; since the correlation functions of the bath decay rapidly, the Lieb-Robinson bound~\cite{Lieb1972,Haah2023} shows that the dissipation term in Eq.~(\ref{eq:RedfieldSchr}) is quasi-local and that its locality is determined by the product of the relaxation time of the correlation function of the bath and the propagation speed in the system~\cite{PhysRevB.111.184311}.

Substituting Eq.~(\ref{eq:adecomp}) into $A_\nu(-s)$ in Eq.~(\ref{eq:RedfieldSchr}), the Redfield equation can be rewritten in terms of the annihilation and creation operators of energy eigenmodes as follows:
\begin{equation}
  \label{eq:singleRF}
  \begin{split}
    \frac{d}{dt}\rho=&-i[H_\mathrm{S},\rho]\\
    &+\sum_jJ_{\mathrm{int},j}^2\sum_{m,n}\left\{\left[V_{mj}V_{nj}^*\Gamma_{11}(\omega_m)(c_m\rho c_n^\dagger-\rho c_n^\dagger c_m)\right.\right.\\
    &\left.\left. +V_{mj}^*V_{nj}\Gamma_{22}(\omega_m)(c_m^\dagger\rho c_n-\rho c_n c_m^\dagger)\right]+\mathrm{H.c.}\right\}.
  \end{split}
\end{equation}
We note that Eq.~(\ref{eq:singleRF}) contains terms in which the creation and annihilation operators of different modes $m,n$ act on the density matrix from the left and right like $c_m\rho c_n^\dag$.
These terms generate quantum coherence in the energy eigenbasis.

The Davies equation can be obtained by applying the secular approximation under which the $m\neq n$ terms in Eq.~(\ref{eq:singleRF}) are ignored as follows~\cite{Davies1974}:
\begin{equation}
  \label{eq:singleDavies}
  \begin{split}
    \frac{d}{dt}\rho=&-i[H_\mathrm{S}+H_\mathrm{LS},\rho]\\
    &+\sum_{j,m}J_{\mathrm{int},j}^2|V_{mj}|^2\gamma_{11}(\omega_m)(c_m\rho c_m^\dag -\{c_m^\dag c_m,\rho\}/2)\\
    &+\sum_{j,m}J_{\mathrm{int},j}^2|V_{mj}|^2\gamma_{22}(\omega_m)(c_m^\dag\rho c_m -\{c_m c_m^\dag,\rho\}/2),
  \end{split}
\end{equation}
where $H_\mathrm{LS}$ is the Lamb-shift Hamiltonian given by $H_\mathrm{LS}=\sum_{j,m}J_{\mathrm{int},j}^2|V_{mj}|^2(\eta_{22}(\omega_m)-\eta_{11}(\omega_m))c_m^\dag c_m$.
By using the KMS condition ~(\ref{eq:KMScond}) of the power spectrum functions, the steady state of Eq.~(\ref{eq:singleDavies}) is shown to be the Gibbs state $\rho_{G}\propto\prod_m\exp[-\beta(\omega_m-\mu)c_m^\dag c_m]$.
It can also be verified that Eq.~(\ref{eq:singleDavies}) satisfies the detailed balance condition,
by showing the following three conditions: (i) \(\rho_G\) commutes with the Lamb-shift Hamiltonian and the Hamiltonian of the system~\cite{Alicki1976,10.1002/9780470142578.ch2,Fagnola2008,fagnola2009two};
(ii) each jump operator removes or adds a definite amount of energy, i.e., the jump operators 
\(
L_m^{(1)}=\sqrt{\sum_jJ_{\mathrm{int},j}^2|V_{mj}|^2\gamma_{11}(\omega_m)} c_m
\)
and
\(
  L_m^{(2)}= \sqrt{\sum_jJ_{\mathrm{int},j}^2|V_{mj}|^2\gamma_{22}(\omega_m)} c_m^\dagger
\)
satisfy
\(\rho_GL_m^{(1)}=e^{\beta(\omega_m-\mu)}L_m^{(1)}\rho_G\) and
\(\rho_GL_m^{(2)}=e^{-\beta(\omega_m-\mu)}L_m^{(2)}\rho_G\);
(iii) the adjoint of the jump operator $L_m^{(1)}$ is also included in the set of the jump operators as
\(e^{-\beta(\omega_m-\mu)/2}L_m^{(1)\dag}=L_m^{(1)}\), which represents the microscopic reversibility of the quantum jumps.

Physically, the secular approximation corresponds to the coarse-graining of the dynamics over time intervals longer than $|\omega_m-\omega_n|^{-1}$ for all the pairs of $m$ and $n$ so that only terms oscillating at the same frequency $\omega_m$ in the interaction picture survive.
For such time coarse-graining to be valid, the time interval must be much shorter than the time scale of the relaxation caused by the system-bath interaction, which requires a sufficiently weak dissipation rate such that $\gamma_{\mu\nu}(\omega)\ll |\omega_m-\omega_n|$.
Thus, the secular approximation breaks down for a large system size because the difference $|\omega_m-\omega_n|$ of the frequencies decreases as the system size increases.

\paragraph{Main result.}
Here we show our main result.
We consider the case where an identical bath is coupled to each site with the same coupling strength $J_{\mathrm{int},j}=J_\mathrm{int}$.
The Hamiltonians of the system, the baths, and the system-bath couplings are given by Eqs.~(\ref{eq:H_system}), (\ref{eq:H_bath}), and (\ref{eq:interaction}).
In this setting, we show that, without any approximation, the Redfield equation~(\ref{eq:RedfieldSchr}) coincides with the Davies equation,
\begin{equation}
  \label{eq:bulkDavies}
  \begin{split}
    \frac{d}{dt}\rho=&-i[H_\mathrm{S},\rho]\\
    &-i\left[J_\mathrm{int}^2\sum_m(\eta_{22}(\omega_m)-\eta_{11}(\omega_m))c_m^\dag c_m,\rho\right]\\
    &+J_\mathrm{int}^2\sum_m\gamma_{11}(\omega_m)(c_m\rho c_m^\dag -\{c_m^\dag c_m,\rho\}/2)\\
    &+J_\mathrm{int}^2\sum_m\gamma_{22}(\omega_m)(c_m^\dag\rho c_m -\{c_m c_m^\dag,\rho\}/2),
  \end{split}
\end{equation}
which means that the Davies equation emerges through a mechanism different from the secular approximation.

Since our derivation does not rely on any time coarse-graining, it does not require the assumption of extremely weak dissipation as in the secular approximation.
Furthermore, the equivalence between the Redfield and Davies equations shows that the quasi-locality of the Redfield equation and the detailed balance condition of the Davies equation are compatible.
In addition, it also shows that the Redfield equation, which does not generally satisfy complete positivity, is completely positive in this case.
We note that a recent study by Nicacio and Koide~\cite{PhysRevE.110.054116} shows the complete positivity and the relaxation to thermal equilibrium of a quadratic master equation given phenomenologically from the canonical quantization of the generalized Brownian motion, while our study gives a microscopic proof of the complete positivity of the Redfield equation based on the Born-Markov approximation starting from the Hamiltonian of the total system.

\paragraph{Derivation.}
When the baths are coupled to each site with the same coupling strength $J_{\mathrm{int},j}=J_\mathrm{int}$, the Redfield equation~(\ref{eq:singleRF}) is rewritten as
\begin{align}
  \label{eq:bulkRF}
  \begin{split}
    &\frac{d}{dt}\rho=-i[H_\mathrm{S},\rho]\\
    &+J_\mathrm{int}^2\sum_{m,n}\left\{\left[\left(\sum_jV_{mj}V_{nj}^*\right)\Gamma_{11}(\omega_m)(c_m\rho c_n^\dagger-\rho c_n^\dagger c_m)\right.\right.\\
    &\left.\left. +\left(\sum_jV_{mj}^*V_{nj}\right)\Gamma_{22}(\omega_m)(c_m^\dagger\rho c_n-\rho c_n c_m^\dagger)\right]+\mathrm{H.c.}\right\}.
  \end{split}
\end{align}
Since $\sum_j V_{mj}V_{nj}^*=\delta_{mn}$ follows from the unitarity of $V$, the $m\neq n$ terms in Eq.~(\ref{eq:bulkRF}) vanish.
Thus, we obtain Eq.~(\ref{eq:bulkDavies}).
Physically, the present derivation indicates that quantum coherences generated by the $m\neq n$ terms by each bath cancel out due to the unitarity of $V$, which leads to the Davies equation.

We note that the derivation can hold for some special cases even if the condition $J_{\mathrm{int},j}=J_{\mathrm{int}}$ is relaxed.
For example, let us consider an \(N\)-site chain with nearest-neighbor hopping and periodic boundary conditions,
where the unitary matrix $V$ is given by $V_{mj}=e^{-imj/N}/\sqrt{N}$.
Let $p$ be a divisor of $N$, and let $J_{\mathrm{int},j}=1$ for sites with $j\equiv 0~(\mathrm{mod}~p)$ and $J_{\mathrm{int},j}=0$ for the other $j$.
Then, $m\neq n$ terms vanish as
$\sum_jJ_{\mathrm{int},j}^2V_{mj}V_{nj}^*=\frac{1}{N}\sum_{j^\prime=1}^{N/p}e^{-i(m-n)pj^\prime/N}=\frac{1}{p}\delta_{mn}$ holds.
The key mechanism here is again the cancellation of quantum coherences generated by each bath.

\paragraph{Time-dependent Hamiltonian.}
In quantum thermodynamics, a time-dependent Hamiltonian is often used to describe an explicit work source.
Here, we extend our result to the case with a time-dependent Hamiltonian $H_\mathrm{S}(t) = \sum_{ij} h_{ij}(t)\,a_i^\dag a_j$,
and derive an instantaneous Davies equation that satisfies the detailed balance condition for the Hamiltonian at each instant if the time scale of the variation of the Hamiltonian is much longer than the relaxation time of the bath.

The Redfield equation for the case with a time-dependent Hamiltonian is expressed in the same form as Eq.~(\ref{eq:RedfieldSchr}), if the operator $A(-s)$ in Eq.~(\ref{eq:RedfieldSchr}) is replaced by
$A_\nu(t,-s) = U^\dag(t,-s)\,A_\nu\,U(t,-s)$, where
$U(t,s) = \mathcal{T}\exp\!\Bigl[-i\int_t^{t+s}H_{\mathrm{S}}(u)\,du\Bigr]$ is the time-evolution operator.
If the time scale $T$ of the variation of the Hamiltonian $H_{\mathrm{S}}(t)$ is much slower than the correlation time \(\tau_B\) of the bath, we can approximate $A_\nu(t,-s)\simeq e^{-isH_\mathrm{S}(t)}\,A_\nu\,e^{isH_\mathrm{S}(t)}$.
Introducing annihilation (creation) operators \(c_{t,m}~(c_{t,m}^\dag)\) of instantaneous energy eigenmodes at time \(t\), such that
\(
H_{\mathrm{S}}(t) = \sum_m \omega_m(t)\,c_{t,m}^\dag c_{t,m},
\)
we obtain, in the same way as the derivation of Eq.~(\ref{eq:bulkDavies}), the following instantaneous Davies equation:
\begin{equation}
  \label{eq:bulkDavies_timedependent}
  \begin{split}
    \frac{d}{dt}\rho &= -i\bigl[H_\mathrm{S}(t),\rho\bigr]\\
    & -i\Bigl[J_\mathrm{int}^2\sum_m\bigl(\eta_{22}(\omega_m(t))-\eta_{11}(\omega_m(t))\bigr)\,c_{t,m}^\dag c_{t,m},\rho\Bigr]\\
    & +J_\mathrm{int}^2\sum_m \gamma_{11}(\omega_m(t))\bigl(c_{t,m}\rho\,c_{t,m}^\dag -\tfrac12\{c_{t,m}^\dag c_{t,m},\rho\}\bigr)\\
    & +J_\mathrm{int}^2\sum_m \gamma_{22}(\omega_m(t))\bigl(c_{t,m}^\dag\rho\,c_{t,m} -\tfrac12\{c_{t,m} c_{t,m}^\dag,\rho\}\bigr).
  \end{split}
\end{equation}

In the conventional derivation of the instantaneous Davies equation by using the secular approximation~\cite{DiMeglio2024timedependent}, we have to assume that the rate of the variation of the system Hamiltonian is slower than the coarse-graining time scale, i.e.,\ an extremely slow variation of the Hamiltonian with $T\gg|\omega_m-\omega_n|^{-1}$ is required.
In contrast, here we only assume the separation $T\gg \tau_B$ of the time scales. Thus, our derivation is applicable to the case where the variation of the Hamiltonian is not extremely slow, which has crucial advantages in applications to quantum thermodynamics.

\paragraph{Nonlinear dissipation.}
The exact equivalence between the Redfield and Davies equations does not hold in general quantum many-body systems.
However, the key mechanism, i.e., cancellation of quantum coherence generated by each bath, may be valid beyond quadratic systems with linear dissipation.
Here we show numerical observations for a system subject to nonlinear dissipation.

Let us assume that the system is coupled to bosonic baths by the interaction Hamiltonian
\(
  H_{\mathrm{SB},j}=J_{\mathrm{int}}\,
  a_{j}^{\dagger}a_{j}\!
  \otimes\!\int dk\,(d_{k}^{(j)}+d_{k}^{(j)\dagger}).
\)
We find that, when the system Hamiltonian is chaotic, the Davies equation remains accurate even for system sizes in which the secular approximation breaks down.
Since the system-bath interaction conserves the particle number in the system, we focus on the single-particle sector below.

We consider two types of system Hamiltonians.
The first one is the Dirac-SYK2 model, whose hopping matrix \(h=(h_{ij})\) is drawn from the Gaussian unitary ensemble of random Hermitian matrices, such that the disorder average of $|h_{ij}|^2$ obeys $\overline{|h_{ij}|^2}=J^2/N$.
The second one is the three-dimensional Anderson model of disordered fermions on a cubic lattice with side length $L$ and system size $N=L^3$, whose Hamiltonian is given by
\begin{equation}
      h_{ij}=\begin{cases}
        w_i\in [-W,W]~(\text{uniformly random, if~}i=j)\\
        J~(\text{if~$i,j$ are nearest neighbor sites})\\
        0~(\text{otherwise}).
      \end{cases}
    \end{equation}
The Dirac-SYK2 model satisfies the single-particle eigenstate-thermalization hypothesis~\cite{PhysRevB.104.214203} and therefore exhibits signatures of quantum chaos.
The Anderson model shows the delocalization-localization transition and an associated chaotic-nonchaotic transition at $W/J \simeq 8.25$~\cite{doi:10.7566/JPSJ.87.094703}.

We assume Ohmic baths with $D(\omega)=|\omega| e^{-|\omega|/\omega_c}$, and then the power spectrum function $\gamma(\omega)$ is given as 
\begin{equation}
  \gamma(\omega)=
  \begin{cases}
    J_\mathrm{int}^2 f_{\beta,\mu,-}(\omega)|\omega| e^{-|\omega|/\omega_c}~~(\omega<0),\\
    J_\mathrm{int}^2 (1+f_{\beta,\mu,-}(\omega))|\omega| e^{-|\omega|/\omega_c}~~(\omega>0).
  \end{cases}
\end{equation}
In the numerical simulation, we neglect the small $\eta(\omega)$ for simplicity.
\begin{figure}
  \centering
  \includegraphics[width=8.6 cm]{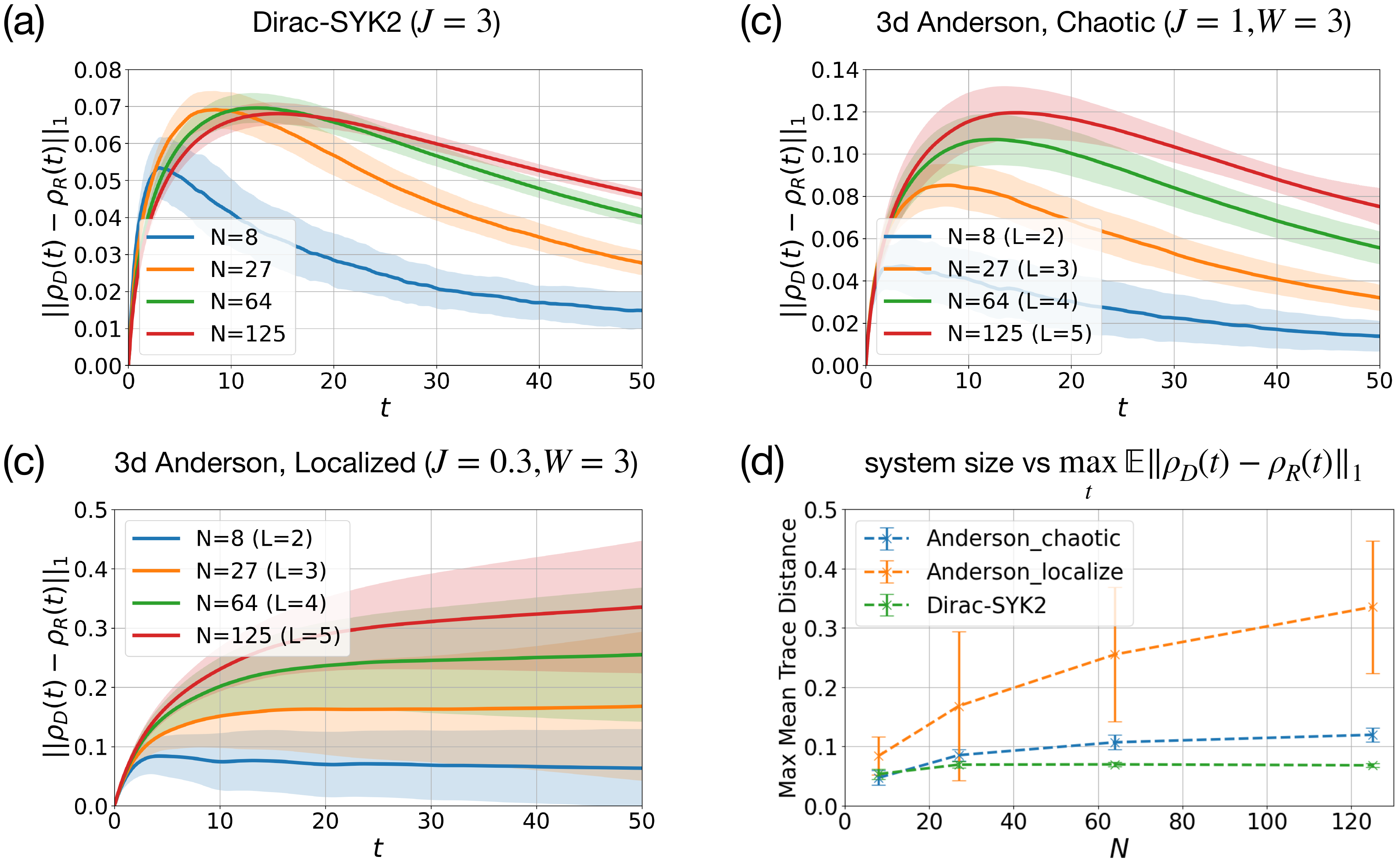}
  \caption{(a-c) Trace distance $\|\rho_D(t)-\rho_R(t)\|_1$ between the density operators obeying the Davies and Redfield equations.
  The solid curves represent the mean values of 100 samples for each system size, and the shadows represent the standard deviation.
  (d) The maximum mean values of $\|\rho_D(t)-\rho_R(t)\|_1$ plotted against the system size. Error bars represent the standard deviations.
  In the chaotic cases, errors are nearly independent of the system size, while in the localized case they increase as the system size increases.
  Other parameters not written in the figure are set to be $J_\mathrm{int}=0.2,\omega_c=10,\beta=5,\mu=0$.
  The initial state is $\rho_D(0)=\rho_R(0)=a_1^\dag\ket{0}\bra{0}a_1$, where $\ket{0}$ is the vacuum state.}
  \label{fig:results}
\end{figure}

We numerically calculate the time evolution of the density operators $\rho_D(t)$ and $\rho_R(t)$ described by the Davies and Redfield equations and compare the two dynamics using the trace distance $\|\rho_D(t)-\rho_R(t)\|_1$ for 100 samples (see Fig.~\ref{fig:results}).
For both the Dirac-SYK\(2\) model and the three-dimensional Anderson model in the chaotic regime, the trace distance $\|\rho_D(t)-\rho_R(t)\|_1$ is nearly independent of the system size.
By contrast, in the localization regime of the Anderson model it grows as the system size increases.
These results suggest that the Davies equation still provides accurate description of the dynamics even when level spacings become small and the secular approximation breaks down.
In general, the validity of the Davies equation would depend on properties of the system Hamiltonian.
In the End Matter, we discuss a possible mechanism of the cancellation of quantum coherence generated by each bath in terms of the chaotic behavior of the system.

\paragraph{Conclusion.}
We have shown that the dynamics of Markovian open quadratic systems coupled to identical baths are exactly described by the Davies equation.
This result indicates that the Redfield equation for those systems already fulfills complete positivity and the detailed balance without the secular approximation.
To the best of our knowledge, this is the first microscopic model where the quasi-locality of dissipation is exactly reconciled with the detailed balance condition embodied by the Davies equation.
Our results are also extended to the case with a time-dependent system Hamiltonian, giving an instantaneous Davies equation valid for drives that are slow compared to the time scale of the baths.
Thus, the derived Davies equation can be applied to a wide range of setups that require locality and thermodynamic consistency.
For example, the present formalism is useful to simulate nonequilibrium dynamics of driven quantum materials coupled to heat baths~\cite{PhysRevB.99.214302,Sato_2019,Sato_2020}.

The mechanism behind our derivation suggests a new possible derivation of the Davies equation for generic many-body systems, when quantum coherences produced by multiple baths cancel each other out.
We have presented numerical results that support this mechanism in the single-particle sector of quadratic systems undergoing nonlinear dissipation.
It merits further study to clarify whether this mechanism works for generic many-body systems.

K.S. was supported by KAKENHI Grant No.~JP23KJ0730 from the Japan Society for the Promotion of Science (JSPS) and FoPM, a WINGS Program, the University of Tokyo.
M.N. was supported by KAKENHI Grant No.~JP20K14383 and No.~JP24K16989 from the JSPS.
T.M. was supported by KAKENHI Grant No.~JP21H05185 from the JSPS and PRESTO Grant No.~JPMJPR2259 from the Japan Science and Technology Agency (JST).

\bibliography{main}
\bibliographystyle{abbrv}
\bibliographystyle{unsrt}

\appendix
\paragraph{Argument for generic many-body systems.}
We discuss a possible mechanism that makes the Davies equation valid in generic quantum many-body systems.
Let \(H_{\mathrm{S}}\) and \(H_{\mathrm{B},j}\) denote the Hamiltonians of a many-body system and a bath coupled to the system at site $j$.
We assume that the system is coupled to identical baths at each site.
Without the loss of generality, the system-bath coupling is written in the form
\begin{equation}
  H_{\mathrm{SB}}= \sum_{j=1}^{N} A_{j}\otimes B_{j},
\end{equation}
where $A_j$ ($B_j$) is an operator of the system (the bath) at site $j$ and $N$ denotes the number of sites.
The correlation function of the bath is given by
\(C(t)=\tr[ B_{j}(t)B_{j}\rho_{\mathrm{B},j}]\), whose one-sided Fourier
transform is denoted by \(\Gamma(\omega)\); by assumption, $C(t)$ and $\Gamma(\omega)$ do not depend on $j$.

Let \(\{ \ket{n}\}\) be energy eigenstates of \(H_{\mathrm{S}}\):
\(H_{\mathrm{S}}\ket{n}=E_n\ket{n}\).
We expand $A_j$ as
\(A_{j}=\sum_{mn} A^{(j)}_{mn}\ket{m}\!\bra{n}\).
The Redfield master equation can then be written as
\begin{equation}
  \label{eq:general}
  \begin{split}
      &\frac{d}{dt}\rho=-i[H_\mathrm{S}+H_\mathrm{LS},\rho]\\
      &+\sum_{k,l,m,n}(\Gamma(E_k-E_l)+\Gamma^{*}(E_m-E_n))\left(\sum_jA_{lk}^{(j)}A_{mn}^{(j)}\right)\\
      &\times \left(\ket{l}\bra{k}\rho\ket{m}\bra{n}-\frac{1}{2}\{\ket{m}\bra{n}\ket{l}\bra{k},\rho\}\right).
    \end{split}
\end{equation}
where the Lamb-shift Hamiltonian is given by
\begin{equation}
  \label{eq:lambshift}
  \begin{split}
    H_\mathrm{LS}=&\sum_{k,l,m,n}\frac{\Gamma(E_k-E_l)-\Gamma^*(E_m-E_n)}{2i}\\
    &\times \left(\sum_jA_{lk}^{(j)}A_{mn}^{(j)}\right)\ket{m}\bra{n}\ket{l}\bra{k}
  \end{split}
\end{equation}

Terms satisfying \(E_k-E_l=E_m-E_n\) in the right-hand side of Eq.~(\ref{eq:general}) constitute the Davies equation.
If energy-level spacings are not degenerate, these terms appear when (i) \(k=m,\;l=n\) or (ii) \(k=l,\;m=n\).
For case (i), $A^{(j)}_{lk}A^{(j)}_{kl}=|A^{(j)}_{kl}|^2$ is nonnegative and thus
\(
  \sum_{j}|A^{(j)}_{kl}|^2
\)
scales as \(O(N)\) with the number of baths (which is equal to the system size).
The similar scaling holds for case (ii).

All other terms with \(E_k-E_l\neq E_m-E_n\) constitute the difference between the Redfield and Davies equations and generate quantum coherence in the energy eigenbasis.
If \(H_{\mathrm{S}}\) is chaotic, i.e., it satisfies the eigenstate thermalization hypothesis (ETH)~\cite{vonNeumann1929Ergodensatz,PhysRevA.43.2046,Srednicki_1999}, we can treat \(A^{(j)}_{lk}\) as independent random complex numbers~\cite{Srednicki_1999}.
If we assume that $A^{(j)}_{lk}A^{(j)}_{mn}$ is a random number independent of $j$, the summation over \(j\) in the right-hand side of Eq.~(\ref{eq:general}) then leads to the cancellation of the quantum coherence generated by the baths at each site, and the non-secular terms scale subextensively in \(N\).
The similar argument can hold for the diagonal and nondiagonal elements of the Lamb-shift Hamiltonian Eq.~(\ref{eq:lambshift}), where the diagonal part of $H_\mathrm{LS}$ commuting with the system Hamiltonian scales as $O(N)$ while the nondiagonal elements scale subextensively.

This argument can be applied to the case of the nonlinear dissipation discussed in the main text.
By taking $\ket{n}=c_n^\dag\ket{0}$ as a basis for the single-particle sector,
the matrix elements $A^{(j)}_{mn}=\bra{m}A^{(j)}\ket{n}$ behave as random variables if the system Hamiltonian satisfies the single-particle ETH~\cite{PhysRevB.104.214203}.
Hence the cancellation mechanism discussed above applies to the case of nonlinear dissipation if the system Hamiltonian is in the chaotic regime.

The argument here implies that, even in generic many-body systems, coherences generated by different baths can cancel out each other,
which suggests that the Davies master equation could remain valid in generic many-body systems even if the secular approximation breaks down.
However, a comprehensive study is needed to establish this mechanism for generic many-body systems, which we leave as future work.

\end{document}